# Thermodynamic Limits of Solar Cells with Non-ideal Optical Response

M. Ryyan Khan, Peter Bermel, and Muhammad A. Alam

Electrical and Computer Engineering Department, Purdue University,
West Lafayette, IN-47906, United States.

***Abstract* — The Shockley-Queisser (S-Q) theory defines the thermodynamic upper limits for Jsc, Voc, FF, and efficiency of a solar cell. The classical calculation assumes an abrupt onset of absorption at the band-edge, perfect absorption for all energies above the bandgap, and absence of non-radiative recombination. These assumptions are never satisfied for any practical solar cell. In this paper, we explain how the S-Q limits are redefined in the presence of the non-ideal optical effects, and we provide closed-form analytical expressions for the new limits for Jsc, Voc, and FF. Remarkably, these new limits can be achieved to a very high degree, even with significantly imperfect materials.**

## I. Introduction

Recent developments in the photovoltaic (PV) technology has resulted in highly efficient cells operating close to the Shockley-Queisser (S-Q) limit [1]. The thermodynamic limits to the various performance matrices such as short circuit current $J_{SC}$, open circuit voltage $V_{OC}$, fill-factor $FF$, and efficiency $\eta_{max}$, have been extensively analyzed in the literature [2–5]. However, these analyses do not consider the intrinsic non-ideal optical responses of the absorber material associated with finite film thickness or weak absorption at the band edge. There have been considerable recent effort to develop advanced optical designs to improve cell efficiency[6–8]. It is, therefore, important to quantify the effects of the non-ideal optical response on the thermodynamic limits of PV performance matrices and establish new limits, if any, due to intrinsic constraints.

In this paper, we consider the non-ideal optical effects for deriving the thermodynamics limits for solar cells. We account for two aspects of this non-ideal response in practical cells: incomplete absorption near the band edge and non-radiative recombination, or less than unity external fluorescence efficiency (EFE) $\eta_{ext}$. The EFE is defined as the fraction of the total recombination that contributes to external radiation from the solar cell. We analytically derive the new 'S-Q limits' for PV performance matrices, i.e., $J_{SC}$, $V_{OC}$, $\eta_{max}$, and $FF$ including the optical non-idealities discussed above. The analytical expressions are in excellent agreement with the corresponding numerical model. Our results imply that, it is possible to operate close to this new S-Q limit despite imperfect absorption; *further optical design aiming towards complete absorptance (e.g., optical black hole* [9]*) would yield negligible improvement in cell efficiency*. Also, low $\eta_{ext}$ can degrade PV efficiency by reducing $V_{OC}$. The conclusions apply and are accurate for broad-range of bandgaps, i.e., $1 < E_g < 2.5$ eV. These predictions from our analytical calculations are also supported by the study based on the well-known PV material, gallium arsenide (GaAs). Finally, we discuss the practical implications of low $\eta_{ext}$ to illustrate the scope for improvement in PV efficiency.

## II. Analytical Expressions For PV Parameters

We assume that sunlight is incident onto the solar cell with a very small solid angle ($\Omega_S \sim 6.8\times10^{-5}$), see Fig. 1(a). An anti-reflection coating (ARC, the blue layer in Fig. 1(a)) suppresses reflection from the top surface. Only a fraction of the light, $A(E)$, is absorbed due to imperfect light trapping. The unabsorbed light $(1-A(E))$ bounces out of the solar cell. If non-radiative recombination is absent (i.e., EFE, $\eta_{ext}=1$), the sum of carriers extracted and photons emitted from the solar cell must equal the number of absorbed incident photons for all voltage $V$. These emitted photons are distributed over a much larger solid angle ($\Omega_D \sim 2\pi$, yellow hemisphere in Fig. 1(a)) compared to that of the incident rays.

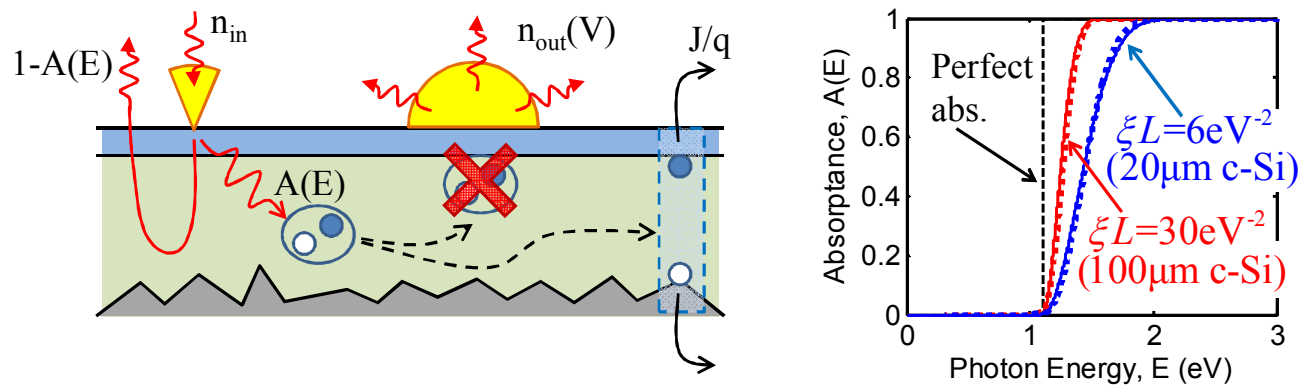

(a) (b)

Fig. 1. (a) Schematic outline of the PV operation (see text for details). (b) Absorptance as a function of energy (at $E_g = 1.1\text{eV}$) for $\xi L = 6/\text{eV}^2$ (blue line) and $30/\text{eV}^2$ (red line). The absorptance for $\xi L = 30/\text{eV}^2$ (red line) matches very well with the absorptance of 100μm c-Si (red dotted line). Correspondingly, $\xi L = 6/\text{eV}^2$ fits 20μm c-Si absorptance (blue solid and dotted lines). The black dashed line shows perfect absorption. Also see Fig. 5 for the absorptance of GaAs.

The empirically fitted absorptance (for energies above the bandgap: $E \ge E_g$) of the described solar cell is approximated as follows [10] (also see part 2, chapter 5 in [11]):

$$A(E) = 1 - e^{-\xi L (E-E_g)^2}. \tag{1}$$

We assume no absorption for energies below the bandgap ($E < E_g$). Here, $\xi L$ denotes the combined effect of the material parameter ($\xi$ is related to the absorption coefficient) and the effective absorption path, $L_{eff}$ (which includes the effect of light trapping). For example, absorptance of $L_{eff} \sim 100\mu\text{m}$ in c-Si ($E_g = 1.1\text{eV}$) is $A_{Si}(E) = 1 - \exp(\alpha(E)L_{eff})$, where $\alpha(E)$ is the absorption coefficient of c-Si [12]. This absorptance spectrum can be approximately fitted by (1) using $\xi L \sim 30/\text{eV}^2$ and $E_g = 1.1\text{eV}$. $A_{Si}(E)$ and the fitted spectrum are shown in Fig. 1(b) as red dotted and solid lines respectively. Direct bandgap materials (e.g., GaAs $E_g \sim 1.42\text{eV}$) have much higher absorption coefficients near the band edge. For instance, absorptance for $L_{eff} \sim 1\mu\text{m}$, GaAs can be approximated with $\xi L \sim 10^3/\text{eV}^2$ and $E_g \sim 1.42\text{eV}$. Fig. 1(b) illustrates the absorptance profile $A(E)$ for two different $\xi L$. Note that, the form for $A(E)$ in (1) is based on the theory of absorption [10] in indirect band-gap materials (e.g., c-Si), therefore it shows good fits for such materials. However, this formula can also be used (approximately) for direct band-gap materials, so long the $L_{eff}$ is sufficiently large to allow moderately good absorption in these films. For example, Eq. (1) fits reasonably well for absorptance spectra for GaAs films thicker than 1-2 $\mu\text{m}$, a dimension typical of practical GaAs solar cells. This length scale is also appropriate for the semi-classical calculations used in this paper.

In the following, we derive expressions for short circuit current $J_{SC}$, open circuit voltage $V_{OC}$, fill factor $FF$, and efficiency $\eta_{max}$, considering imperfect absorption near the band edge and degraded external fluorescence efficiency ($\eta_{ext}$). Note that, the external fluorescence efficiency is given by the fraction of recombination which is emitted as radiation from the solar cell [13].

### *A. The J-V relationship*

The photon flux per energy radiated from a blackbody with chemical potential $\mu$ at temperature $T$ is given by [4],

$$n_{rad}(E,T,\mu,\Omega) = \frac{2\Omega}{c^2h^3}\frac{E^2}{e^{(E-\mu)/kT}-1}. \tag{2}$$

Here, $\Omega$ is the solid angle covered by the concerning radiation, $c$ is the speed of light in free space, $h$ is the Planck's constant, and $k$ is the Boltzmann constant. The photon flux per energy incident on the solar cell from the sun is taken to be $n_{in}(E) = n_{rad}(E,T_S,\mu=0,\Omega_S)$. This idealized spectrum resembles the standard extraterrestrial (AM0) solar spectrum. Now, the emission from the solar cell operating at voltage $V$ would be characterized by $n_{out}(E,qV) = n_{rad}(E,T_D,\mu=qV,\Omega_D)$. In our calculations we assume $T_S = 6000$ K and $T_D = 300$ K [4].

The principle of 'Detailed balance' ensures that the number of carriers extracted from the solar cell equals the absorbed and emitted photons. Thus, the net current is given by:

$$J(V) = q\int_{Eg}^{\infty}\left(n_{in}(E) - \frac{n_{out}(E,qV)}{\eta_{ext}}\right)\times A(E)dE \tag{3}$$

$$\begin{aligned} &= q\int_{Eg}^{\infty}\left(n_{in}(E) - \frac{n_{out}(E,qV)}{\eta_{ext}}\right)dE \\ &\quad - q\int_{Eg}^{\infty}\left(n_{in}(E) - \frac{n_{out}(E,qV)}{\eta_{ext}}\right)\times e^{-\xi L(E-E_g)^2}dE \\ &= J_L(V) - J_{NA}(V,\xi L). \end{aligned} \tag{4}$$

Here, $\eta_{ext}$ is the external fluorescence efficiency. Equation (1) was used in the second line in the above expressions. Note that, $J_L$ is the current corresponding to perfect absorption above $E_g$. The effect of imperfect absorption is reflected in the second term $J_{NA}$. We find that the $J-V$ relationship can be obtained analytically by using Boltzmann approximation, as follows:

$$J(V) = q\Omega_S\tilde{\gamma}(T_S)e^{-Eg/kT_S} - \frac{q\Omega_D\tilde{\gamma}(T_D)e^{-Eg/kT_D}e^{qV/kT_D}}{\eta_{ext}}. \tag{5}$$

The last term in Eq. (5) highlights the importance of improving $\eta_{ext}$ for highly efficient solar cell. Here, $\tilde{\gamma}(T)$ is the contribution from the 3D photonic density of states, and,

$$\tilde{\gamma}(T) \equiv \gamma_{SQ}(T) - \gamma_{NA}(T,\xi L) \tag{5A}$$

$$\gamma_{SQ}(T) = \frac{2kT}{c^2h^3}\left(E_g^2 + 2kTE_g + 2k^2T^2\right) \tag{5B}$$

$$\gamma_{NA}(T,\xi L) \approx \frac{2kT}{c^2h^3}E_g^2\,\Delta(\xi L). \tag{5C}$$

Here, we define an absorption non-ideality term as,

$$\Delta(\xi L) = \frac{1-e^{-(2/\sqrt{\pi})\sqrt{\xi L}\,kT}}{(2/\sqrt{\pi})\sqrt{\xi L}\,kT}. \tag{5D}$$

Remarkably, the effect of incomplete absorption is accounted for by a simple multiplicative factor, $\Delta(\xi L)$. For the small values of $\xi L$ associated with poor absorbers, the parameter $\Delta(\xi L) \to 1$. For very strong absorption (high $\xi L$), the parameter $\Delta(\xi L) \to 0$. Note that, by allowing $\Delta(\xi L) \to 0$ and $\eta_{ext} = 1$, we return to the S-Q limits exactly as presented in Ref. 4.

### *B. Short circuit current and open circuit voltage*

Short circuit current can be obtained from (5) by setting $V = 0$:

$$J_{SC} = q\Omega_S\tilde{\gamma}(T_S)e^{-Eg/kT_S} - q\frac{\Omega_D}{\eta_{ext}}\tilde{\gamma}(T_D)e^{-Eg/kT_D}. \tag{6}$$

The open circuit voltage $V_{OC}$ is obtained by setting $J(V_{OC}) = 0$ in (5):

$$\begin{aligned} qV_{OC} = E_g\left(1 - \frac{T_D}{T_S}\right) - kT_D\ln\left(\frac{\Omega_D}{\Omega_S}\right) - kT_D\ln\left(\frac{\gamma_{SQ}(T_D)}{\gamma_{SQ}(T_S)}\right) \\ - kT_D\ln\left(\frac{1}{\eta_{ext}}\right) + kT_D\ln\left(\frac{1}{1-\Delta}\right). \end{aligned} \tag{7}$$

This is important generalization of the expressions for $V_{OC}$ given in Refs. 4, 5, and 8. The logarithmic suppression of $V_{OC}$ due to poor $\eta_{ext}$ is clearly indicated. Physically, low $\eta_{ext}$ reflects reduction in carrier buildup (and hence lowered $V_{OC}$) due to non-radiative recombination. Also, note that the last term reflects change in $V_{OC}$ due to incomplete absorption $(\Delta > 0)$. Interestingly, $V_{OC}$ *increases from its S-Q value due to imperfect absorption at the band edge* (which causes effective widening of the optical bandgap). However, this increase in counterbalanced by a reduction in $J_{SC}$, so that the overall efficiency remain below the classical S-Q limit. See Sec. III for additional discussion on this topic.

### *C. Efficiency and Fill-Factor*

The maximum efficiency of a solar cell is written as the ratio of the output power $(J_{opt}V_{opt})$ at optimum operating condition to the incident solar power $P_{in}$:

$$\eta_{max} = \frac{J_{opt}V_{opt}}{P_{in}}. \tag{8}$$

The input solar power is $P_{in} = 12\Omega_S (kT_S)^4/(c^2h^3)$. The optimum operating condition can be derived by maximizing efficiency $(\eta = JV/P_{in})$ with respect to $V$. By setting $d\eta/dV = 0$ we solve for $V$ to find $V_{opt}$:

$$qV_{opt} = E_g\left(1-\frac{T_D}{T_S}\right) - kT_D \ln\left(\frac{\Omega_D}{\Omega_S}\right) - kT_D \ln\left(\frac{1}{\eta_{ext}}\right) + kT_D \ln\left(\frac{1}{1-\Delta}\right). \tag{9}$$

Note that, as seen from (7) and (9), both $V_{OC}$ and $V_{opt}$ are shifted by the same amount due to the non-ideal optical effects considered here (effect of both $\eta_{ext} < 1$ and $\Delta(\xi L) > 0$). The optimum current is found from $J(V = V_{opt})$ in (5):

$$J_{opt} = q\Omega_S \tilde{\gamma}(T_S) e^{-Eg/kT_S} - q\frac{\Omega_D}{\eta_{ext}}\tilde{\gamma}(T_D) e^{-Eg/kT_D} e^{qV_{opt}/kT_D}$$
$$= q\Omega_S e^{-Eg/kT_S}\left(\tilde{\gamma}(T_S) - \tilde{\gamma}(T_D)\frac{1}{1-\Delta}\right). \tag{10}$$

The second line utilizes (9) for the final expression of $J_{opt}$. As explained earlier, $\Delta \rightarrow (1-\grave{o})$ for smaller values of $\xi L$. Hence $1/(1-\Delta)$ is very high for low $\xi L$, saturating rapidly to a low value with increasing $\xi L$. This means that while $V_{opt}$ increases for low absorption, $J_{opt}$ decreases rapidly. Indeed, since the decrease in $J_{opt}$ (decrease $\sim 1/(1-\Delta)$) is faster compared to increase in $V_{opt}$ (increase $\sim \ln[1/(1-\Delta)]$) with lowered absorption, together they degrade $\eta_{max}(\propto J_{opt}V_{opt})$ below the S-Q limit. The fill-factor

$$FF = \frac{J_{opt}V_{opt}}{J_{sc}V_{OC}} \tag{11}$$

can be obtained from previously derived expressions for $J_{opt}$, $V_{opt}$, $J_{sc}$, and $V_{OC}$. We find that all our analytically derived values from (6)-(11) are accurate within 7% to the numerical thermodynamic calculation results for practical solar cell bandgap range of $1\text{eV} < E_g < 2\text{eV}$. The results of this comparison are summarized in Fig. 2, which can be used to translate the analytical results to accurate numerical results with appropriate scaling.

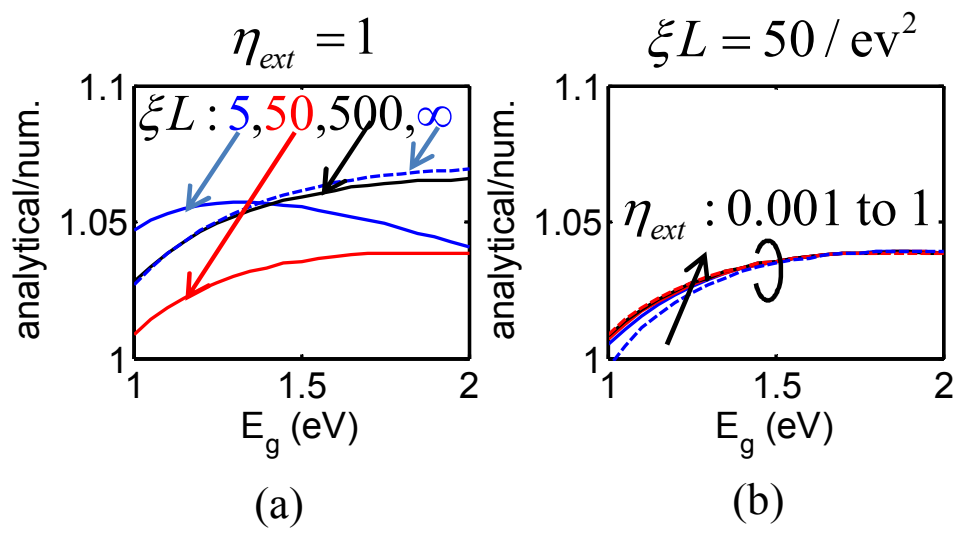

Fig. 2. Accuracy of the analytical model in terms of ratio of efficiencies calculated analytically to those found numerically: (a) for various $\xi L$ values and, (b) for various $\eta_{ext}$. The errors for various $\xi L$ and $\eta_{ext}$ are within 7% of the numerically calculated values for $1\text{eV} < E_g < 2\text{eV}$.

## III. Discussion

### *A. Insights from the analytical relationships*

Fig. 3 shows that poor absorption with low $\xi L$ (blue curves) degrades PV performance in terms of $J_{SC}$ and efficiency $\eta_{max}$, but counter-intuitively it improves $V_{OC}$ slightly compared to higher absorption (red curves). Recall that $V_{OC}$ is determined by the emission spectrum $(n_{out} \times A(E))$ of the solar cell; the weighting of the emission by $A(E)$ shifts the emission peak away from $E_g$ towards higher energies. This effective widening of the optical bandgap increases $V_{OC}$. Note that this effective widening of the optical bandgap essentially will shift the S-Q efficiency vs. $E_g$ curve to the left yielding slightly improved PV performance for the smaller bandgap solar cells (where $E_g < 1.35\text{eV}$). This concept could be useful for PV materials with bandgap lower than the S-Q optimum ($E_g^{SQ-opt} \approx 1.35\text{eV}$).

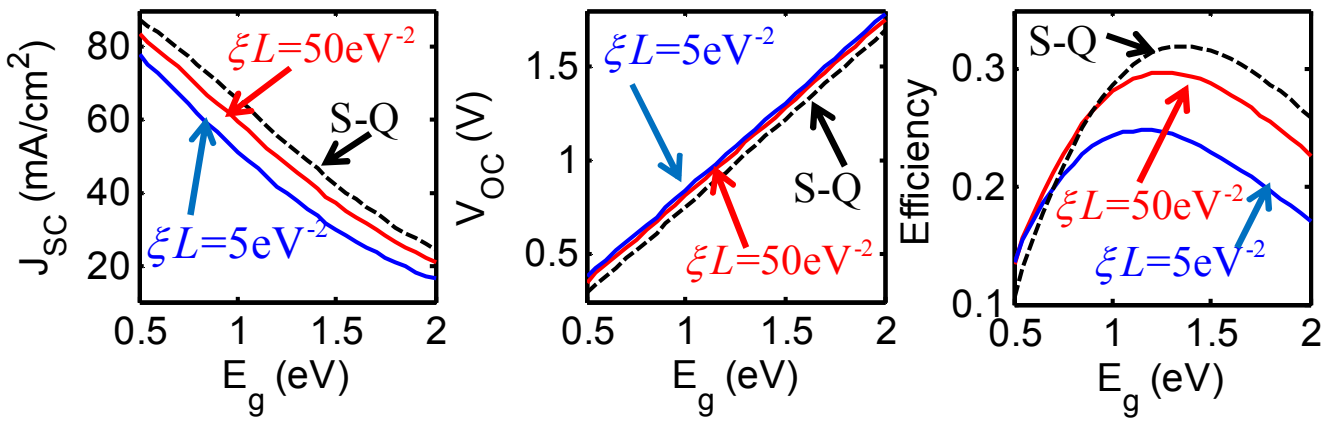

Fig. 3. Analytically derived solar cell parameter ($J_{SC}$, $V_{OC}$, and $\eta_{max}$) limits at $\xi L = 5/\text{eV}^2$ (blue line) and $50/\text{eV}^2$ (red line). The S-Q limit is shown as the black dashed lines (corresponds to $\xi L \rightarrow \infty$). For these results we assume $\eta_{ext} = 1$.

What would be the implication of dramatically improving absorption (parameterized by $\xi L$) for a cell that originally had very poor absorption? Consider a GaAs solar cell with $E_g = 1.42\text{eV}$ and $\eta_{ext} = 1$. The blue solid curves in Fig. 4 show that the effect of $\xi L$ saturates very quickly after $\xi L \sim 1000/\text{eV}^2$. This translates to $L_{eff} \sim 1\mu\text{m}$ for GaAs. After this critical point, $J_{SC}$ increases very slowly, $V_{OC}$ decreases very slowly, and, $\eta_{max}$ saturates close to the S-Q limit. This indicates that the efficiency would essentially saturate for a GaAs solar cell with finite absorptance. Although $J_{SC}$ has not reached its maximum possible value at this $\xi L$ point, the increased $V_{OC}$ compensates for the lowered $J_{SC}$ to yield an $\eta_{max}$ approaching S-Q value from below due to imperfect absorption.

Figure 4 also demonstrates the effect of reduced external fluorescence efficiency $\eta_{ext}$. Only a small fraction of the recombination in indirect bandgap materials occurs radiatively. As seen in Fig. 4, for $\eta_{ext} = 0.2$ [14] the efficiency limit is lowered even for perfect absorption. As expected, $\eta_{ext}$ does not affect $J_{SC}$. This is because $\eta_{ext}$ only modifies the emission process and thus alters the $V_{OC}$. The non-radiative recombination ( $\eta_{ext} < 1$ ) reduces the $V_{OC}$ which in turn degrades $\eta_{max}$.

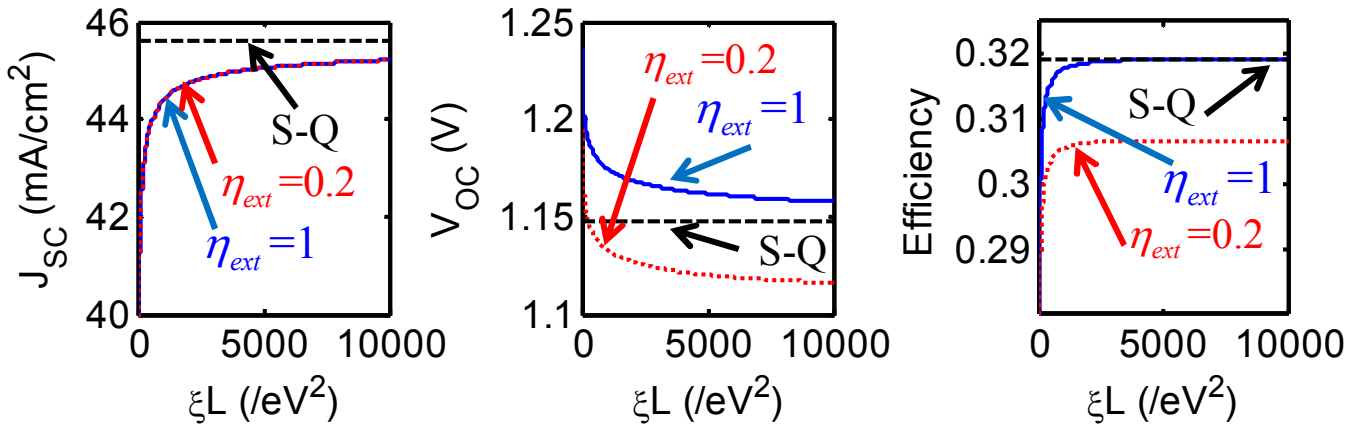

Fig. 4. Effect of varying $\xi L$ on solar cell parameter ( $J_{SC}$, $V_{OC}$, and $\eta_{max}$ ) limits at $\eta_{ext} = 1$ (blue line) and $\eta_{ext} = 0.2$ (red dotted line). The black dashed lines give the S-Q limit.

### B. *Practical implications: incomplete absorption and Urbach Tails*

From a practical perspective, let us investigate incomplete absorption in GaAs. The absorption coefficient $\alpha$ of GaAs shown in Fig. 5(a) has been obtained from [12]. Although $E_g = 1.42$ eV for GaAs, the absorption in the Urbach tail [15]—indicated by the shaded region in the inset plot of Fig. 5(a) -- makes $\alpha$ non-zero even for $E < E_g$. The corresponding absorptance, $A(E) = 1 - \exp(\alpha(E)L)$, of a GaAs film of thickness $L$ is shown in Fig. 5(b). For smaller $L$ values ( $1\mu m$ and $0.1\mu m$, represented by blue and red lines, respectively), the exponentially small $\alpha$ in the Urbach tail ensures that the contribution from these states to overall absorption is negligible. However, for a very thick film ( $L = 10^3 \mu m$ represented by the black line), the relative contribution by the Urbach tail increases significantly, especially for the photons with $1.42 > E > 1.3$eV. This can be thought of as effective lowering of the optical bandgap below 1.42 eV.

We do expect some deviations in the PV performance parameters for real materials (e.g., GaAs) from the estimates obtained using our simplified analytical absorption model. However, the predicted trends are surprisingly robust—which we will show for practical case of GaAs solar cells. Fig. 6 shows the PV parameters ( $J_{SC}$, $V_{OC}$ and efficiency) as a function of GaAs film thickness $L$. The J-V relationships and the corresponding PV parameters in this case have been calculated numerically based on (3), with AM1.5 illumination. For $\eta_{ext} = 1$ (blue line, Fig. 6), initially $J_{SC}$ rises very quickly with $L$, however, the rate of increase decreases sharply once $L > 3\mu m$, see Fig. 6(a). We also observe continually decreasing $V_{OC}$ as a function of $L$, as discussed earlier. Finally, we observe that the efficiency $\eta_{max}$ quickly rises for very thin solar cells, but then saturates for $L > 3\mu m$. Beyond this point, gradual increase in $J_{SC}$ is counterbalanced by a gradual decrease in $V_{OC}$ resulting in a saturated $\eta_{max}$ versus $L$ relationship. Note that, the $J_{SC}$ gain shown in the shaded region of Fig. 6(a) is contributed by the Urbach tail, however, the absorption in the Urbach tail reduces the effective optical bandgap, yielding in a degraded $V_{OC}$. The increased $J_{SC}$ is compromised by a corresponding degraded $V_{OC}$, keeping the efficiency approximately constant in these $L$ values.

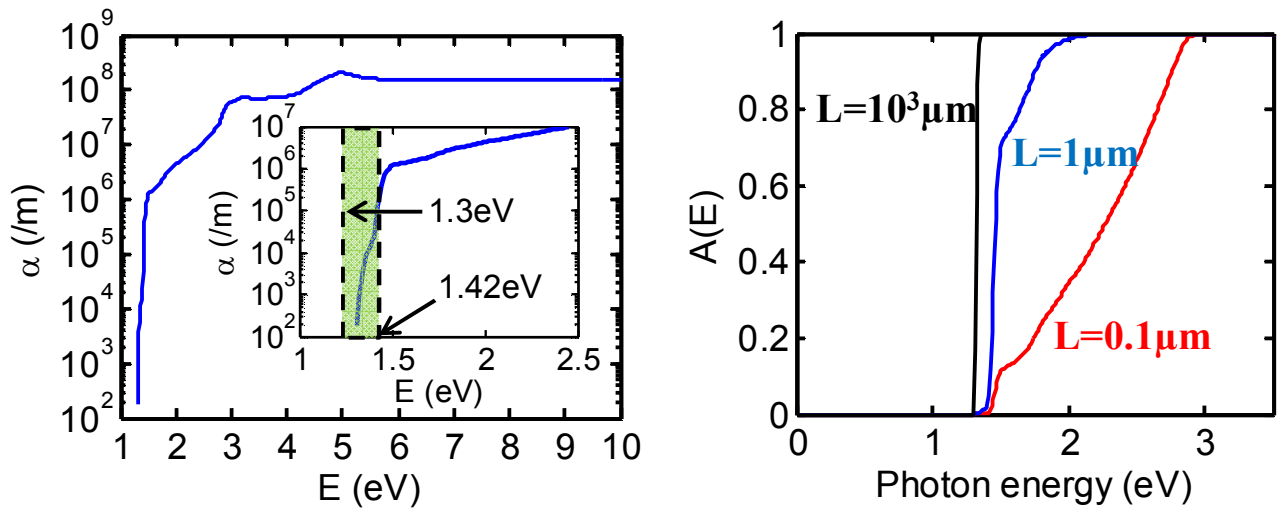

(a) (b)

Fig. 5. (a) Absorption coefficient spectrum of GaAs. The shaded region in the inset plot shows the Urbach tail. (b) The absorptance spectrum of a GaAs film of thickness $0.1\mu m$ (red), $1\mu m$ (blue), and $10^3 \mu m$ (black).

The effect of non-radiative recombination is essentially the same as discussed earlier in Sec. III.A. Reduced $\eta_{ext} (= 0.2)$ decreases $V_{OC}$ without having any effect on $J_{SC}$ —thus degrading the efficiency (see red curves in Fig. 6).

In short, our analysis of GaAs predicts that we do not require very high light harvesting to reach the ultimate (S-Q) solar cell efficiency. A GaAs solar cell of thickness $\sim 3\mu m$ can very closely approach the performance limit associated with this material bandgap.

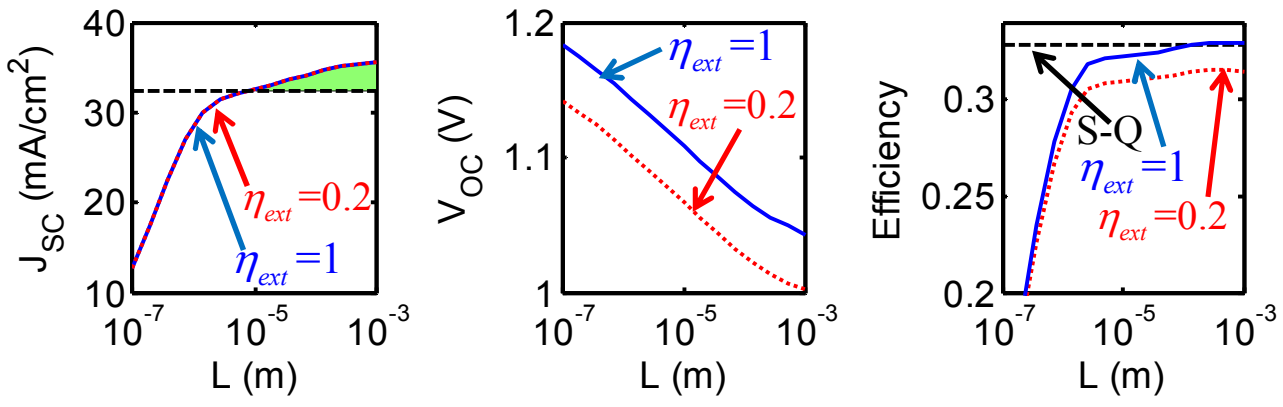

(a) (b) (c)

Fig. 6. Effect of varying $L$ on GaAs solar cell parameter ( $J_{SC}$, $V_{OC}$, and $\eta_{max}$ ) limits at $\eta_{ext} = 1$ (blue line) and $\eta_{ext} = 0.2$ (red dotted line). The black dashed lines in (a), (c) is the S-Q efficiency limit.

### C. Practical implications: non-radiative recombination

Equation (7) suggests that $V_{OC}$ is reduced by $kT_D \ln(1/\eta_{ext})$ in presence of non-radiative recombination, i.e., for $\eta_{ext} < 1$. This provides us with the opportunity for possible improvement in $V_{OC}$ by increasing $\eta_{ext}$ —Table I estimates this value for some well-known solar cells [14]. To stress the importance of improved $\eta_{ext}$, it is obvious that the increase in $V_{OC}$ by ~70mV in GaAs (Alta) cells compared to GaAs (ISE) cells can be explained exclusively by the enhancement in $\eta_{ext}$. This improved $V_{OC}$ yielded the highest efficiency GaAs solar cell by Alta Devices [16]. Room for improvement by enhancing $\eta_{ext}$ in Si-UNSW and Si-SPWR devices is remarkably close (see Table I). However, these Si solar cells have different $V_{OC}$ values which can be attributed to other losses in the devices. CIGS (along with most other thin-film materials) still has the greatest potential for improvement.

TABLE I

THE $V_{OC}$ AND $\eta_{ext}$ VALUES FOR VARIOUS AVAILABLE SOLAR CELL DEVICES [14] ARE SHOWN HERE. THE ROOM FOR $V_{OC}$ IMPROVEMENT ASSOCIATED WITH $\eta_{ext}$ IS ESTIMATED IN THE RIGHT-MOST COLUMN

| Device | $V_{OC}$ (mV) | $\eta_{ext}$ (%) | $\Delta V_{OC}^{(EFE)} = kT_D \ln(1/\eta_{ext})$ (mV) |
|---|---|---|---|
| Si UNSW | 706 | 0.57 | 129 |
| Si SPWR | 721 | 0.56 | 130 |
| GaAs Alta | 1107 | 22.5 | 37 |
| GaAs ISE | 1030 | 1.26 | 109 |
| CIGS(NREL) | 713 | 0.057 | 187 |

## IV. CONCLUSIONS

We have derived the thermodynamics performance limit of solar cells in the presence of imperfect absorption and non-ideal external fluorescence efficiency. The expressions illustrate, in a compact analytical form, the effects of imperfect optical absorption and non-radiative recombination. We find that approaching S-Q limit does not require *perfect* absorption and therefore, the need for perfect optical design can be relaxed considerably. Numerical analysis based GaAs further reinforced this conclusion. Finally, solar cells with low $\eta_{ext}$ have room for improvement in $V_{OC}$ and thus in efficiency. Opto-electronic design aiming towards devices with predominantly radiative recombination would be of prime interest for this purpose.

## ACKNOWLEDGEMENTS

This material is based upon work supported as part of the Center for Re-Defining Photovoltaic Efficiency Through Molecule Scale Control, an Energy Frontier Research Center funded by the U.S. Department of Energy, Office of Science, Office of Basic Energy Sciences under Award Number DE-SC0001085. The computational resources for this work were provided by the Network of Computational Nanotechnology under NSF Award EEC-0228390. This was also supported by the Bay Area PV Consortium, a Department of Energy project with Prime Award number DE-EE0004946.

## REFERENCES

[1] W. Shockley and H. J. Queisser, "Detailed Balance Limit of Efficiency of p-n Junction Solar Cells," *J. Appl. Phys.*, vol. 32, no. 3, p. 510, 1961.

[2] A. Luque and S. Hegedus, Eds., *Handbook of Photovoltaic Science and Engineering, Second Edition*. 2011.

[3] J. Nelson, *The physics of solar cells*. London: Imperial College Press, 2004.

[4] L. C. Hirst and N. J. Ekins-Daukes, "Fundamental losses in solar cells," *Prog. Photovolt: Res. Appl.*, vol. 19, no. 3, pp. 286–293, 2011.

[5] M. A. Alam and M. R. Khan, "Fundamentals of PV Efficiency Interpreted by a Two-Level Model," *arXiv:1205.6652*, May 2012.

[6] J. N. Munday, D. M. Callahan, and H. A. Atwater, "Light trapping beyond the 4n2 limit in thin waveguides," *Applied Physics Letters*, vol. 100, no. 12, p. 121121–121121–4, Mar. 2012.

[7] D. M. Callahan, J. N. Munday, and H. A. Atwater, "Solar Cell Light Trapping beyond the Ray Optic Limit," *Nano Lett.*, vol. 12, no. 1, pp. 214–218, 2011.

[8] A. Polman and H. A. Atwater, "Photonic design principles for ultrahigh-efficiency photovoltaics," *Nat Mater*, vol. 11, no. 3, pp. 174–177, Mar. 2012.

[9] E. E. Narimanov and A. V. Kildishev, "Optical black hole: Broadband omnidirectional light absorber," *Applied Physics Letters*, vol. 95, no. 4, p. 041106, 2009.

[10] S. Datta, *Quantum Phenomena*. Addison-Wesley, 1989.

[11] M. S. Dresselhaus, "6.732 SOLID STATE PHYSICS." [Online]. Available: http://web.mit.edu/course/6/6.732/www/texts.html..

[12] "Refractive Index Database – Table of Refractive Index Values for Thin Film Thickness Measurement." [Online]. Available: http://www.filmetrics.com/refractive-index-database. [Accessed: 09-Jul-2012].

[13] O. D. Miller, E. Yablonovitch, and S. R. Kurtz, "Intense Internal and External Fluorescence as Solar Cells Approach the Shockley-Queisser Efficiency Limit," *arXiv:1106.1603v3*, Jun. 2011.

[14] M. A. Green, "Radiative efficiency of state-of-the-art photovoltaic cells," *Progress in Photovoltaics: Research and Applications*, 2011.

[15] F. Urbach, "The Long-Wavelength Edge of Photographic Sensitivity and of the Electronic Absorption of Solids," *Phys. Rev.*, vol. 92, no. 5, p. 1324–1324, Dec. 1953.

[16] "Alta Devices: Finding a Solar Solution - Technology Review," *Technology Review*. [Online]. Available: http://www.technologyreview.com/energy/39649/. [Accessed: 27-Mar-2012].